\RequirePackage{fix-cm}
\documentclass[smallextended]{svjour3}       
\smartqed  
\usepackage{graphicx}
\usepackage{amsmath}  
\usepackage{amsfonts} 
\usepackage{graphicx} 
\usepackage{color}
\usepackage{comment}
\usepackage{tabularx}
\newcommand{\gam}{{\nu}}

\newcommand{\be}{\begin{equation}}
\newcommand{\ee}{\end{equation}}

\newcommand{\ep}{\epsilon}

\newcommand{\ex}{\kappa}

\newcommand{\SB}{Stefan-Boltzmann }
\newcommand{\IDOS}{integrated density of states }
\newcommand{\DOS}{density of states }

\newcommand{\red}{\text}

\begin{document}

\journalname{published in Foundations of Physics 48, p.395 (2018)}

\title{Generalized Stefan-Boltzmann law
}


\author{Gilles Montambaux
}


\institute{G. Montambaux \at
              Laboratoire de Physique des Solides, CNRS, Universit\'e Paris-Sud, Universit\'e Paris-Saclay, 91405- Orsay, France \\
              \email{gilles.montambaux@u-psud.fr}           
}

\date{Received: Aug. 3, 2017 / Accepted: March 13, 2018}

\maketitle

\begin{abstract}
We reconsider the thermodynamic derivation by L. Boltzmann of the Stefan law and we generalize it for various different physical systems {\it whose chemical potential  vanishes}. Being only based on classical arguments, therefore independent of the quantum statistics, this derivation applies as well to the saturated Bose gas in various geometries as to "compensated" Fermi gas near a neutrality point, such as a gas of Weyl Fermions. It unifies in the same framework the thermodynamics of many different bosonic or fermionic non-interacting gases which were until now described in completely different contexts.
\end{abstract}

\section{Introduction}

 In 1879, J.  Stefan found experimentally that the total power emitted by a black body scales as the fourth power of the temperature.\cite{stefan1879}
  This law was derived theoretically by L. Boltzmann in 1884,\cite{boltzmann1884} without any knowledge on the {\it spectral repartition} of the black body emission
   which was discovered later by M. Planck.\cite{planck1901} Nowadays,  the now well-established expression of the \SB law with its correct prefactor\cite{remark2}
\be u(T)= {\pi^2 k_B^4 \over 15 \hbar^3  c^3}  \  {  T^4} \ , \label{SBlawc} \ee
where $u=U/V$ is the energy per volume unit, is simply deduced
 in most textbooks from the Planck law integrated over frequencies.  It is however interesting to recall that the {\it total} emission can be derived from  purely classical and thermodynamical grounds, as Boltzmann did.
 The main two ingredients for this derivation are the second law of thermodynamics and the result  obtained by J.C. Maxwell\cite{maxwell1873} that the  radiation  exerts a pressure
 $p(T)$ proportional to the total energy density, $p(T)= {1 \over 3} u(T)$ (This pressure is now understood as the pressure of {\it massless} particles which are the photons).  The  $1/3$ factor, obtained by Boltzmann, results  from angular averaging in $3D$ and is indeed the inverse of the space dimensionality.

  In this paper, we show how   the \SB (SB) law can be generalized to any gas of   free particles, and we stress the importance of a third ingredient for its derivation, a {\it zero chemical potential}, as it was implicit in the Boltzmann argument.
 In particular, we show how the simple relation between pressure and internal energy for a gas of {\it massive} particles $p(T) = {2 \over 3} u(T)$  leads directly,    {\it on purely classical grounds},
  to the temperature dependence of the total energy of a $3D$  saturated Bose gas, that is an ideal gas below the Bose-Einstein
  condensation temperature.

More generally, we consider  free  gases of particles with  general, including hybrid dispersion relations, and show that the classical argument applies not only to the saturated Bose gas in various geometries (box or harmonic potentials), but also to gases of Fermions in various situations under the condition that the Fermi energy is temperature independent, which is the case for half-filled band systems. Such systems will be called "compensated Fermi gases".
We regroup in the same framework the thermodynamic properties of many different bosonic or fermionic free gases, which until now were described in completely different contexts. The outline of the paper is the following. In next section, we recall the main elements of the Boltzmann derivation, including homogeneity arguments.
In section \ref{sec.Massive}, we generalize it to massive particles and show that this classical derivation accounts for the thermodynamics of the saturated Bose gas. More general spectra are considered in section \ref{sec.General}, where we discuss specifically the case of the saturated Bose gas in various geometries including boxed or harmonic potentials. In section \ref{sec.HybridS}, we consider gases of particles with hybrid spectrum, that is massive in some directions, massless in the other directions, and we present several examples including bosons in mixed boxed-harmonic potential or fermions with  exotic dispersion relation (section \ref{sec.CompensatedF}). For particles which are massive in $d_m$ directions and massless in $d_c$ directions, the generalized Stefan-Bolzmann law may be summarized in the form
\be u(T) \propto    { (mc^2)^{d-\ex}  \over (h c)^d} (k_B T)^{\ex+1} \ ,    \ee
where $d=d_c+d_m$ and $\kappa=d_c+d_m/2$.
We conclude by the presentation of a table with many different physical situations described by this generalized Stefan-Boltzmann law.

\medskip

\section{Stefan-Boltzmann law}

We first briefly recall the derivation of this law as first theoretically derived by Boltzmann.\cite{boltzmann1884}
We start by considering the entropy $S(U,V)$.  The variation of entropy for an infinitesimal transformation is
 \be T d S= dU + p dV =     {\partial U \over   \partial T} d T +  \left(p+ {\partial U \over \partial V} \right) d V  \ . \label{SB1} \ee
Since the entropy is a total differential, the equality  between cross derivatives leads to:
 \be  {\partial p \over \partial T}= {p + u \over T} .  \label{dpuT} \ee
  Inserting the Maxwell-Boltzmann relation $p(T) = {u(T) / 3}$ in  Eq.(\ref{dpuT}) gives immediately

\be {1 \over 4}  {d u \over  u }= {d T \over T} \ , \label{du4T} \ee
leading to the Stefan law $u(T) = K_1 T^4$.
Implicitly, in Eq.(\ref{SB1}) the chemical potential is set to zero, which is essential for the demonstration here.

Of course the correct numerical factor can only be obtained from integration over frequencies of the spectral function $u(\nu,T)$ and requires the famous form of the Planck law discovered later in 1900.\cite{planck1901}
However, the combination of universal constants could be  obtained simply by dimensional arguments, where a quantity having the dimension of an action is needed.
 The Planck constant was not introduced at that time, but the existence of a natural unit for action does not require quantization hypothesis. Indeed such natural unit was already introduced in 1899 by Planck\cite{planck1899}(see Appendix A), so that Eq.(\ref{SBlawc}) without the numerical factor
  can be obtained from purely classical arguments, $h$  being a natural unit of action obtained experimentally\cite{planck1899}(see also \cite{Paul}).

We now turn to the generalization of the SB law to a gas of massive particles with zero chemical potential.

\medskip

\section{Massive particles}
\label{sec.Massive}

Since the unique assumptions for deriving SB law are the thermodynamic  principles and the relation between
 internal energy and pressure, it is interesting
to reconsider now the Boltzmann derivation  in the case of massive particles. However an important remark is in order.
Contrary to photons whose number is
 not fixed, the number of massive particles is in general fixed (in average) and controlled by the chemical potential $\mu$.
  Eq.(\ref{SB1})  should indeed be written as
\be T d S =  d U  + p dV  - \mu d N \ . \label{TSUVN} \ee
Therefore, the  use of  Eq.(\ref{SB1}) implies that  the chemical potential is zero, which is actually the case for a gas of photons, leading to the SB law.
 If one now considers a gas of massive particles, we know from Maxwell-Boltzmann kinetic theory of gases that $p$ and $u$ are  related by $p={2 \over 3} u$. The  Boltzmann derivation still holds
{\it   under the condition that  $\mu=0$}.
 In this case, from  (\ref{dpuT}) one gets, instead of (\ref{du4T}):
 \be {2 \over 5} { d u \over  u }= {d T \over T} \label{du25T} \ee
 so that $u(T)=K_2 T^{5/2}$. A dimensional argument which necessarily involves an action, the Boltzmann constant and a mass (instead of a velocity as before) leads to
\be u(T) \propto  {m^{3/2} \over h^3} (k_B T)^{5/2} \label{SBlawm} \ee
This behavior is actually well-known: this is  the internal energy of the $3D$ saturated Bose gas, that is an ideal gas below the Bose condensation temperature,  or of  a gas  magnon excitations in a $3D$ ferromagnet.
We stress that this expression does not require the knowledge of the thermal repartition of energy levels and no quantum mechanical input: the introduction of $h$ as a unit of action in the prefactor does not imply any quantization. It is therefore quite interesting that the {\it thermodynamic of the saturated gas can be described in the framework of classical physics, with the
unique hypothesis that $\mu=0$}.

More generally, the  Maxwell-Boltzmann kinetic theory of gases can be used for  any dispersion relation between energy and momentum of the form $\ep \propto p^\gam$ in dimension $d$, and  leads to the general relation  $u(T)= \ex p(T)$ with $\ex= d/\gam$ ($\gam=1$ for photons and $\gam=2$ for massive particles). The thermodynamics of such gases is described below.

\medskip

\section{General spectrum}
\label{sec.General}

 Consider  now any gas of particles at zero chemical potential whose relation between pressure and energy is given a linear relation  $u(T)= \ex p(T)$.
Then
Eqs.(\ref{du4T},\ref{du25T})  have to be replaced by\cite{Lima}
\be {1 \over \ex +1} {d u \over u}= { d T \over T}  \ . \ee
Integration of this relation together with dimensionality arguments leads to the connection between energy-pressure relation and the generalized \SB law:

\be   u(T)= \ex p(T)  \implies   \displaystyle u(T) \propto { (mc^2)^{d - \ex} \over (h c )^d} (k_B T)^{\ex +1}   \label{SBlawkappa}\ee
which generalizes Eqs.(\ref{SBlawc},\ref{SBlawm}).

We propose here a quicker, simpler and more general  derivation~:
instead of pressure, which may not be well defined in certain situations that we will consider below, we consider the grand potential $\Omega$, and
suppose a relation of the form $U = - \ex \Omega$ which generalizes the relation $u = \ex p$. Then the thermodynamic relation between the grand potential  and the energy (this is an interesting exercise to show that the following relation is equivalent to the thermodynamic relation $\Omega= U - T S - \mu N$, see Appendix B):
  \be  U = \left({\partial \beta \Omega \over \partial \beta}\right)_{\! \alpha} \label{UbetaOmega} \ee
 where $\alpha = \beta \mu=\mu/(k_B T)$,  together with $U= - \ex \Omega$, leads
immediately  to
 \be \left({\partial \beta U  \over \beta U  }\right)_{\! \alpha} = - \ex {\partial \beta \over \beta} \ . \ee
Integration of this relation, combined with dimensional arguments, gives
\be U(T) = F(\alpha)  \left( {L \over h c} \right)^d  (mc^2)^{d-\ex}  (k_B T)^{\ex+1} \ , \label{UTFmc}  \ee
 where $F(\alpha)$ is a multiplicative integration function. Here no assumption is made on the structure of $F(\alpha)$ (which is actually an integral of Bose
 or Fermi factors\cite{Fgf}). The generalized Boltzmann law (\ref{SBlawkappa}) follows immediately when the chemical potential vanishes.   Finally since the number of particles $N(T)$ is given by $N(T)= - \left.{\partial \beta  \Omega \over \partial \alpha} \right|_\beta$, we find quite generally that
 \be U(T)= \kappa {F(\alpha) \over F'(\alpha)}\,  N(T) k_B T  \ . \label{UNkT} \ee

\medskip

\noindent
In the following, we introduce the thermal de Broglie wavelength $\lambda_{dB} =  h/(2 \pi m k_B T )^{1/2}$ and another characteristic wavelength for a massless spectrum
$\lambda_W = h c/ k_B T$, that we may call "Wien wavelength" since it separates classical and quantum behaviors. Then,  the total number of particles   can be rewritten as (within a multiplicative numerical constant):

\be N(T) =  F'(\alpha)  {V \over \lambda_{dB}^{2(d - \ex)} \lambda_W^{2 \ex - d}}  \label{NTdB}\ee
and the total energy is given by (\ref{UNkT})  which are forms of the generalized SB law (\ref{SBlawkappa}).
The relations (\ref{UNkT},\ref{NTdB}) are general, for any $\alpha$. Then two cases are of interest:
\medskip

$\bullet$ For the {\it classical} ideal gas where the particle number $N$ is {\it fixed} and $U(T)= \kappa N k_B T$, one deduces that $F(\alpha)=e^{\alpha}$, and from (\ref{NTdB}) one finds the relation for the chemical potential
\be \mu = - k_B T \ln {V \over N \lambda_{dB}^{2(d - \ex)} \lambda_W^{2 \ex - d}}  \ee
which generalizes the case of massive particles ($\kappa=d/2$).
\medskip

$\bullet$ When $\alpha=0$, for example for the {\it saturated ideal gas} where the chemical potential is fixed to $0$, we get the generalized Stefan-Boltzmann law. Then the number of particles above the condensate $N(T)$ varies with temperature, and we have the relations
\be U(T)= \kappa N(T) k_B T \qquad , \qquad S(T)= (\kappa+1 ) N(T) k_B  \ .   \ee
\medskip


For more complex dispersion relations as hybrid dispersion relations discussed later in sections \ref{sec.HybridS}, \ref{sec.CompensatedF}, the calculation of pressure from kinetic theory may be  difficult to achieve because of
 complex angular averages. The pressure may even not be properly defined.
 Then we may  obtain the coefficient $\ex$ and deduce the SB law directly from the
energy dependence of the \DOS $\rho(\ep)$, when it is a power law. Here we rather consider  the {\it integrated} \DOS (also called counting function) denoted as $N_{\!<}(\ep)$.
We start with an  \IDOS of the form $N_{\!<}(\ep) = B \ep^{\ex}$, where $B$ has necessarily the dimensions
\be B \propto L^d {(mc^2)^{d - \ex} \over (hc)^d} \ .  \label{B} \ee
 When $\alpha=0$, the total energy $U(T)$ has the
  form

\be
U(T)= \int \ep \rho(\ep) f(\beta  \ep) d\ep   \label{uTpT} \ee
where no assumption is needed on the form on the function $f$ except that it is a function of the single combination
 $\beta \ep$ (which is Wien's displacement law for light). This immediately leads to a power law $T^{\ex+1}$  for the total energy and to the relation $U(T)= - \ex \Omega(T)$ between energy and grand potential.
Therefore, we have the following correspondence
 \be \rho(\ep) \propto \ep^{\ex-1} \  \Longleftrightarrow \  U = - \ex \Omega \ \Longleftrightarrow  \ U \propto T^{\ex+1}  \ . \ee
This is the main result of this paper. It may appear quite natural since the grand potential is actually a double integral of the density of states. It has however a profound origin which is Wien's displacement law which is valid for any gas of {\it zero chemical potential}.
Since it is obtained from purely classical arguments, no assumption is made on the nature of the particles. The argument applies as  well  for Bosons as for  Fermions under the condition that $\mu=0$.
In the following, we review  many different physical systems whose thermodynamics is unified in this unique framework.

\medskip

 Until now, we have only considered a free gas in a box of volume $V$. Particularly important in the context of Bose-Einstein condensation (BEC), is the
 case of a Bose gas trapped in a harmonic
 potential. To avoid any detailed calculations, it is useful to remark that the density of states of massive particles in a harmonic trap is the same
 as   the one of free massless particles. This is of course reminiscent of the wave-particle duality  that the radiation spectrum can be
 described either as an ensemble of resonators or as
 a gas of massless particles. The substitution to be done is
  \be  {\pi c \over L} \rightarrow \omega \quad \text{or} \quad
 {L \over \lambda_W} \rightarrow {k_B T \over 2 \hbar \omega} \label{subst} \ee
and we have  $\ex=d$. Therefore the  \IDOS has the form, for an isotropic harmonic well
 \be N_{\!<}(\ep) \propto \left( {\ep \over \hbar \omega } \right)^d  \ .  \ee
 Again $\hbar$ has been introduced from homogeneity,
 but the scaling dependence of the \DOS does not need quantization. The above argument leads to
 \be N(T) \propto \left( k_B T \over \hbar \omega \right)^d  \qquad , \qquad U(T) \propto N(T) k_B T  \ee

Notice that in this case Maxwell-Boltzmann argument cannot be used directly since volume and pressure are obviously not defined. However, Ref.\cite{harmonicpressure} defines a   "harmonic volume"  ${\cal V} = {1/\omega}^3$, consistent with the above substitution (\ref{subst}), and the conjugated "harmonic pressure" ${\cal P}= - {\partial \Omega /\partial {\cal V}}$.\cite{harmonicpressure}
\medskip

\medskip

\section{Hybrid spectrum}
\label{sec.HybridS}

Up to now, we have only considered situations of either massless or massive
 particles, so that obviously a mass and a velocity cannot enter both in (\ref{SBlawkappa},\ref{UTFmc},\ref{NTdB}). One has respectively either $\ex=d$ or
 $\ex=d/2$, so that
\be N(T)= {V \over   \lambda_W^{d} } \qquad \text{or} \qquad N(T)= {V \over \lambda_{dB}^{d} } \ .
\ee
Interesting is the more general situation of a {\it hybrid} spectrum, massive in $d_m$ directions and massless in $d_c directions$. In this case, the Maxwell relations $u=3p$ for massless particles or $u=3p/2$ for massive particles can be generalized to
 \be u= \left(d_c + {d_m \over 2}\right) p \ . \ee
Quite generally, assume a dispersion relation written in the short dimensionless notation
\be \ep \sim  \sum_j A_j  n_j^{\gam_j}  \label{spectrebetaj} \ee
with $\gam_j=2$ along $d_m$ massive directions and $\gam_j=1$ along $d_c$ massless directions. $A_j$ have the dimensions of an energy and $n_j$ are positive integer numbers.
The \IDOS is still a power $N_{\!<}(\ep) = B \ep^\ex$ with
\be \ex=  \sum_j {1 \over \gam_j} = d_c + {d_m \over 2}  \label{dosdcdm} \ee
and $B$ is given by Eq.(\ref{B}). Therefore the number of particles has the form
\be N(T)\propto {V \over \lambda_{dB}^{d_m}\, \lambda_W^{d_c}} \label{NTdBlambda}\ee
and the expressions for energy and pressure result immediately.

\medskip


For completeness, we consider the hybrid situation of a saturated Bose gas in a potential with a  box (or size $L$) profile
 in $d_m$ directions and an harmonic profile in $d_c$ directions.  The substitution (\ref{subst}) in Eq.(\ref{NTdBlambda}) gives

\be N(T)= \left({L \over \lambda_{dB}}\right)^{d_m}  \left( k_B T \over \hbar \omega \right)^{d_{c}} \ .  \ee

\medskip

\section{Compensated Fermi gas}
\label{sec.CompensatedF}

\begin{figure}[h!]
\begin{center}               \includegraphics[width=8cm]{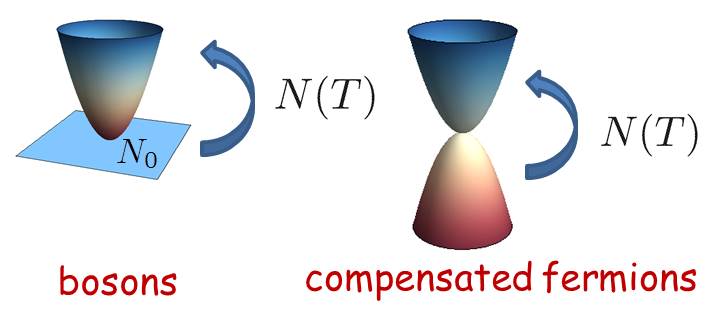}
\end{center}
\caption{Left : a Bose gas having $N_0$ particles in the ground state. At finite temperature, $N(T)$ particules are excited and form the saturated Bose gas,  keeping the chemical potential $\mu=0$, until the Bose-Einstein temperature $T_c$ is reached when $N(T_c)=N_0$. Right: An electron compensated gas transfers $N(T)$ electrons from the valence band to the conduction band, keeping $\mu=0$ when the spectrum is symmetric. There is no transition in this case since the reservoir is infinite. }
\label{fig:bosons-fermions}
\end{figure}

Another situation where the Boltzmann derivation can be applied is the one of a {\it Fermi} gas, under the condition that  $\mu=0$. This peculiar situation occurs when the Fermi level is pinned at a neutrality point with electron-hole symmetry.
This is for example the case for Dirac massless fermions, extensively studied to describe electronic states of graphene at half-filling,
where the Fermi energy lies in the middle of the energy spectrum and the chemical potential {\it is independent
of temperature}. We denote this system as a "compensated Fermi gas". The $T=0K$ filled lower band plays the role of the condensate, and the finite $T$ excited electrons play the role of the saturated gas (Fig. \ref{fig:bosons-fermions}).

 \medskip


Consider first the case of $2D$ Dirac fermions like in graphene, for which $d=2$, $\gam=1$, so that $\ex=2$ (the \DOS is linear). The general relation (\ref{SBlawkappa})
gives immediately
\be u(T) \propto   { (k_B T)^{3} \over (h c)^2} \ee
leading to the well-known specific heat $\propto T^2$ of graphene for the half-filled band.\cite{grapheneRev} Similarly,
 in bilayer graphene, the spectrum is quadratic so that $\gam=2$ and $\ex=1$, and from (\ref{SBlawkappa}) we have
\be u(T) \propto   { m \over h^2} (k_B T)^{2}  \ . \ee

The above arguments can  be applied to the now recently extensively studied Weyl fermions in  semi-metals.\cite{Rao}
Consider   Weyl Fermions in $d$ dimensions. The energy spectrum being linear, one has $\gam=1$, and $\ex=d$, so that
\be u(T) \propto { (k_B T)^{d+1} \over (h c )^d} \ee
The thermodynamics of Weyl Fermions is therefore the same as the one of the black body.

Finally, we remind that for an intrinsic compensated semiconductor in $3D$, in the limit of a vanishing gap, the number of elecrons in the conduction band varies as
\be N(T) \propto {m^{3/2} \over h^3} (k_B T)^{3/2} \ee
which is also a form of the Stefan-Boltzmann law.

\medskip

     \begin{figure}[h!]
\begin{center}               \includegraphics[width=8cm]{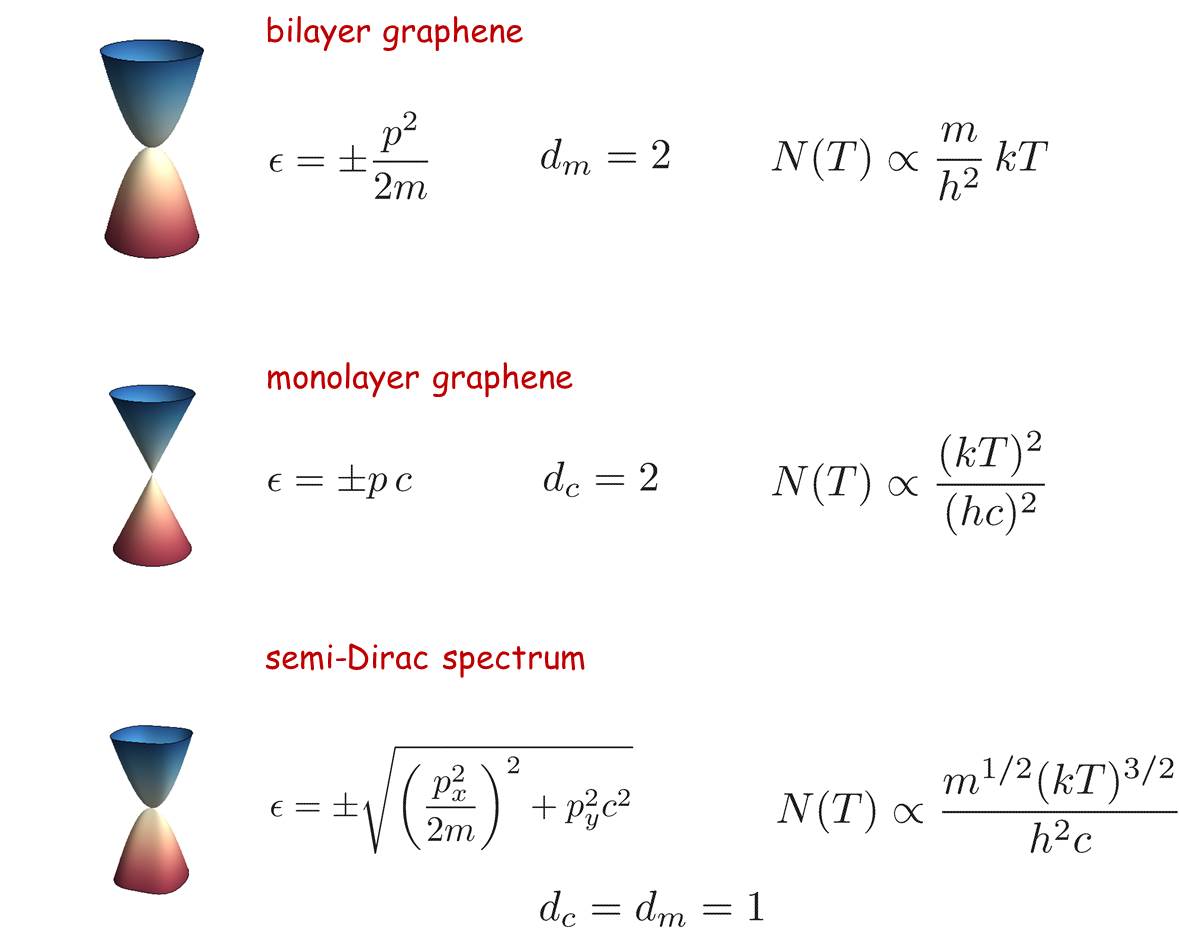}
\end{center}
\caption{Stefan-Boltzmann law for compensated Fermions, massive, massless and hybrid particles}
\label{fig:fermions}
\end{figure}

It has been shown that under appropriate lattice distortion, the two Dirac points characteristic of the graphene electronic spectrum can merge into a single point, where the spectrum stays linear
 in one direction but becomes quadratic along the other direction.\cite{Guinea2008,Montambaux2009} This hybrid point has been called a "semi-Dirac" point.\cite{Pickett} This is an interesting example of an hybrid spectrum characterized by both a mass and a velocity (Fig. \ref{fig:fermions}).

\be \ep= \pm \sqrt{ \left({p_x^2 \over 2 m}\right)^2 + c^2 p_y^2} \ee
whose thermodynamic properties have been studied.\cite{Guinea2008,Montambaux2009}
This is special case of a general dispersion relation of the form
\be \ep \sim \pm  \sqrt{ \sum_j A_j n_j^{2 \gam_j}}  \ee
which leads to the same energy dependence of the density of states as a spectrum of the form (\ref{spectrebetaj}) (see Appendix C), and the energy density if given by Eq.(\ref{SBlawkappa}) where $\ex= d_c+d_m/2= 3/2$.
The thermodynamic properties of a semi-Dirac Fermi spectrum are therefore the same as the saturated Bose gas
  in a mixed boxed-harmonic potential. More generally a gas of Weyl fermions close to a merging point, the spectrum can transform from a full linear dispersion relation to a linear-linear-massive, or a linear-massive-massive, to a full massive spectrum. The corresponding thermodynamics is summarized in Table \ref{tableau}.
\medskip

     \begin{figure}[h!]
\begin{center}               \includegraphics[width=8cm]{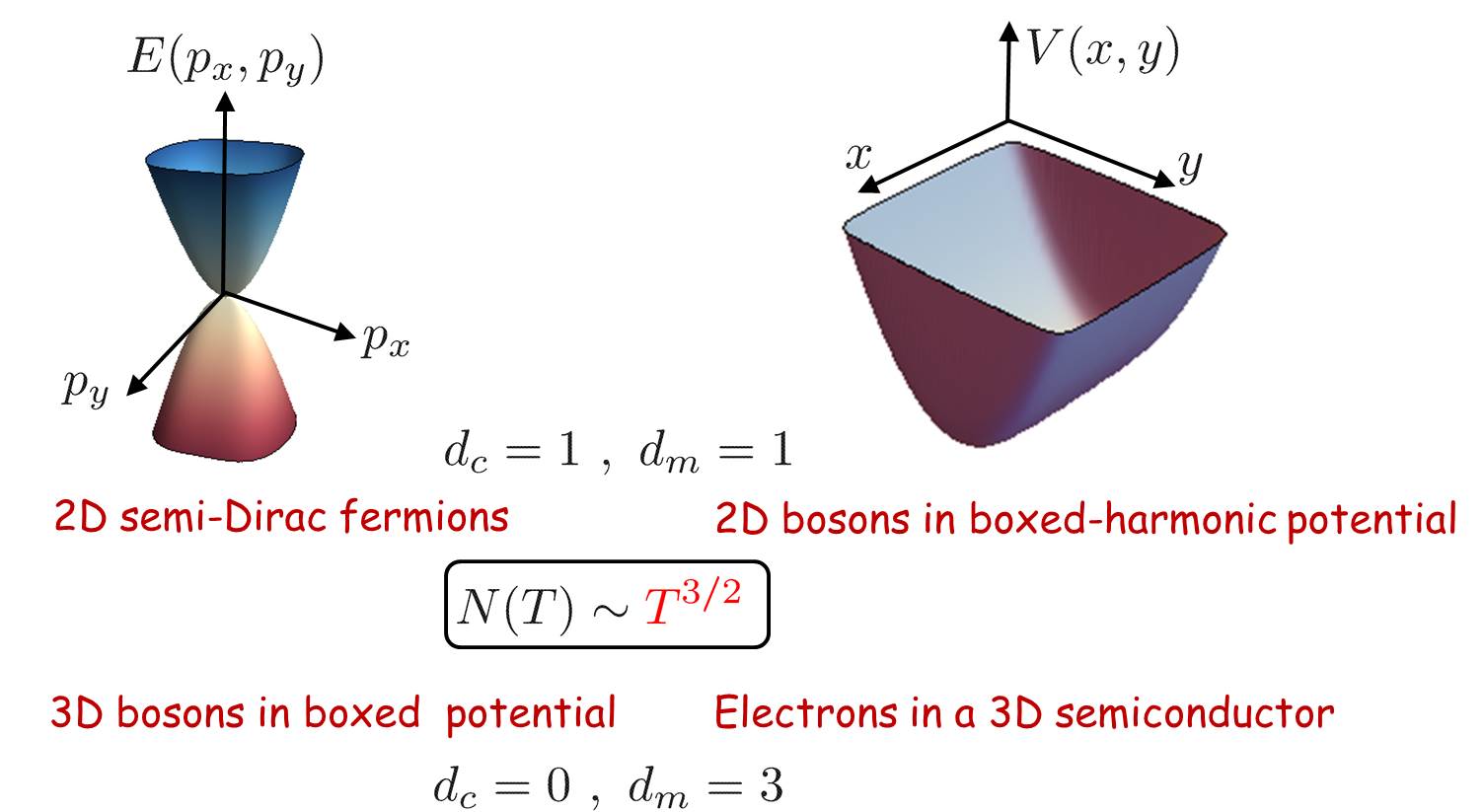}
\end{center}
\caption{Semi-Dirac fermions in $2D$ (let), a Bose gas in a $2D$ harmonic-boxed potentiel  (right ), a $3D$ Bose gas in a boxed potential, or a compensated semiconductor in $3D$ (limit of small gap) obey the same Stefan-Boltzmann law. }
\label{fig:unif}
\end{figure}

\section{Conclusion}

We have generalized the Boltzmann derivation of the \SB law to ideal gases of any kind of particles, under the condition that the chemical potential stays zero.
From the Maxwell-Boltzmann relation between energy and pressure, one simply recovers the thermodynamic properties of a saturated gas of massive particle below the Bose-Einstein condensation temperature.
Then we generalize the derivation to the case of hybrid dispersion relations, massive in some directions, massless in other directions. This situation is achieved in Bose gases confined in
 hybrid boxed-harmonic potentials.
We have also shown that fermionic gases with particle-hole symmmetry behave indeed as saturated gases with fixed chemical potential. This
remark leads to draw similarities between quite different problems like a semi-Dirac Fermi gas and a 2D Bose gas in a mixed boxed-harmonic potential.
 The recently studied physics of Weyl points in 3D is even richer since the dispersion relation can be linear in two directions and massive in a third direction, a situation thermodynamically similar to a 3D Bose gas confined
 in a harmonic trap in two directions and a boxed potential in the third direction.
The unifying picture is summarized in Table \ref{tableau}.

\medskip


\begin{table*}[h!]
\begin{center}
\begin{tabular}{|c|c|c|c|c|c|}
  \hline
   & $\ex=1$ & $\ex=3/2$ & $\ex=2$ & $\ex=5/2$ & $\ex=3$ \\[5pt]
   \hline
  d=2 &    &   &{\red  black body} &  &  \\
    & {\it  massive fermions} & {\it semi-Dirac fermions}  & {\it massless fermions}  &  & \\
  & $\displaystyle N(T) \propto {L \over \lambda_{dB}^2}$ & $\displaystyle  N(T) \propto {S \over \lambda_{dB} \lambda_W}$  &
   $\displaystyle  N(T) \propto  {S \over \lambda_W^2}$ &  & \\[10pt]
\hline
d=2&    & {\red  BEC + $\sqcup$ + $\smile$  } & {\red BEC + $\smile$ + $\smile$ }&  &  \\
 &    & $\displaystyle N(T) \propto{L \over \lambda_{dB}} {k_B T \over \hbar \omega}$   & $\displaystyle N(T) \propto\left({k_B T \over \hbar \omega}\right)^2$ &  &  \\[10pt]
  \hline
  \hline
  d=3 &   & &  & & {\red black body}  \\
 &   &{\red  F-magnons}  &  & & {\red phonons, AF-magnons} \\
 &  & {\it massive fermions}  &  {\it hybrid MML fermions}  & {\it hybrid MLL fermions} &  {\it massless fermions} \\
  &  & $\displaystyle N(T) \propto {V \over \lambda_{dB}^3}$   & $\displaystyle N(T)\propto {V \over \lambda_{dB}^2 \lambda_W}$
    &
$\displaystyle  N(T) \propto {V \over \lambda_{dB} \lambda_W^2}$  &  $\displaystyle N(T) \propto {V \over  \lambda_W^3}$  \\[10pt]
\hline
d=3 & &  {\red BEC  + $\sqcup$ + $\sqcup$ + $\sqcup$ } & {\red   BEC  + $\sqcup$ + $\sqcup$ + $\smile$ }  & {\red BEC  +
 $\sqcup$ + $\smile$ + $\smile$} &  {\red  BEC  + $\smile$ + $\smile$ + $\smile$}\\
& & $\displaystyle N(T) \propto {V \over \lambda_{dB}^3}$ &  $\displaystyle N(T)\propto {S \over \lambda_{dB}^2 }{k_B T \over \hbar \omega}$
 &  $\displaystyle N(T)\propto {L \over \lambda_{dB} }\left({k_B T \over \hbar \omega}\right)^2$ &
 $\displaystyle N(T)\propto  \left({k_B T \over \hbar \omega}\right)^3$ \\[10pt]
  \hline
   $u(T)$   & $T^2$ & $T^{5/2}$  & $T^3$  &  $T^{7/2}$  &  $T^4$   \\[10pt]
  \hline
\end{tabular}
\end{center}
  \caption{This table summarizes the thermodynamic properties of various   massive, massless or hybrid saturated ($\mu=0$) gases. The properties depend both on the space dimensionality $d$ and the proportionality coefficient $\ex$ between internal
energy $u(T)$ and pressure $p(T)$. The symbols $\sqcup$  and $\smile$  denote respectively a box potential or a harmonic
potential in a given direction. The examples of bosons are shown in normal characters, and the example of fermions are denoted in italics. MML denotes a massive-massive-massless spectrum,  MLL   a massive-linear-linear spectrum. One has $N(T) \propto T^\ex$ and $U(T) \propto T^{\ex+1}$.
}
\label{tableau}
\end{table*}

\clearpage

\appendix

\section{Planck's natural unit of action}
\label{AppA}

In his 1899 paper,\cite{planck1899} more than one year before his discovery of the black body law with quantization
 of energy exchange, M. Planck found a natural unit of action from his analysis of the \SB law and
Wien's  law:
  \be u(\nu,T)= { 8 \pi b \over c^3} \nu^3  e^{- a \nu /T} \ee
  where $a$ and $b$ are parameters to be fixed from the experiment.
  Comparing with experimental estimate of the total energy
\be u(T)= \int_0^\infty u(\nu,T) d \nu= { 48 \pi b \over a^4 c^3}\,  T^4 \ee and Wien's displacement law, Planck could extract the parameters $b/a^4$ and $a$, and obtain the values
 $b= 6.885 \, 10^{-34}$ J.s and $a= 0.4818 \, 10^{-10}$ K.s.
This was the first estimate of a unit of action, which was not yet related to any idea of quantization, and which will be denoted later as $h$ in the 1900 discovery of the black body law.
Moreover, in our modern notations, $a= h/k_B$ so that  the obtained value of $a$ leads to an estimate of  $k_B = 1.43 \, 10^{-23}$J/K (that Planck will introduce one year later in his seminal paper).

\section{Thermodynamic relations}

Here we show the equivalence between these to relations

  \be  U = \left({\partial \beta \Omega \over \partial \beta}\right)_{\! \alpha}  \quad \Longleftrightarrow  \quad \Omega=U - TS - \mu N \ee
  One first explicitly write the derivative as
  \be U = \beta   \left({\partial \Omega \over \partial \beta}\right)_{\! \alpha} + \Omega = - T \left({\partial \Omega \over \partial T}\right)_{\! \alpha}+ \Omega \label{der} \ee
  The next step is to rewrite $\left({\partial \Omega \over \partial T}\right)_{\! \alpha}$.
We first write the grand potential $\Omega$ as ($k_B=1$ in this section)
\be \Omega\left(\alpha= {\mu \over  T},T \right) \ . \ee
Consider the derivatives
\begin{align}
\left({\partial \Omega \over \partial \mu}\right)_{\! T} &= {1 \over T} \left({\partial \Omega \over \partial \alpha}\right)_{\! T}  \\
\left({\partial \Omega \over \partial T}\right)_{\! \mu} &= -{\mu \over T^2} \left({\partial \Omega \over \partial \alpha}\right)_{\! T}+
\left({\partial \Omega \over \partial T}\right)_{\! \alpha}
\end{align}
Inserting the first equation into the second one in order to eliminate $\alpha$, we obtain
\be T \left({\partial \Omega \over \partial T}\right)_{\! \alpha}= \mu \left({\partial \Omega \over \partial \mu}\right)_{\! T} + T \left({\partial \Omega \over \partial T}\right)_{\! \mu} = -  \mu N - TS  \label{appB1} \ee
Inserting this last relation in eq. (\ref{der}), we find the expected result.

\section{Exact thermodynamic results}

We give here for completeness the exact expressions for the density of states and thermodynamic quantities of gases of particles in various dimensions and geometries.

\subsection{Integrated density of states}

Here we give complete expressions fot the integrated \DOS for particles with various dispersion relations.

$\bullet$ Free massive particles, $\ep= {p^2 \over 2 m}$ :
\be N_{\!<}(\ep)= C_d \left({L \over 2 \pi}\right)^d \left({2 m \ep \over \hbar^2}\right)^{d/2}  \ .  \label{eq44} \ee
where $C_d= \pi^{d/2}/\Gamma(1+d/2)$ is the unit volume in $d$ dimensions.
\medskip

$\bullet$ Free massless particles, $\ep = |p| c$ :
\be N_{\!<}(\ep)= C_d \left({L \over 2 \pi}\right)^d \left({  \ep \over \hbar c}\right)^{d} \ .  \label{eq45} \ee

$\bullet$ Particles in a isotropic harmonic potential, $\ep= (n_x+n_y+n_z) \hbar \omega$ :
\be N_{\!<}(\ep)= {1 \over d!} \left({  \ep \over \hbar \omega}\right)^{d} \ .  \label{eq46}\ee

$\bullet$ $2D$ semi-Dirac fermions $\ep =\sqrt{\left({p_x^2 \over 2 m}\right)^2 + p_y^2 c^2}$ :\cite{Montambaux2009}
\be N_{\!<}(\ep)=\left({L \over \hbar}\right)^2 C {\sqrt{m} \over   c } \ep^{3/2} \,  \ee
with $C= {\Gamma(1/4)^2 \over 6 \pi^{5/2}}$.
\bigskip

$\bullet$ $2D$ particles in a boxed-harmonic potential, $\ep= {p_x^2 \over 2 m} + n_y \hbar \omega$ :
\be N_{\!<}(\ep)= {2 L \over 3 \pi }  {\sqrt{m} \over \hbar^2  \omega } \ep^{3/2}  \ .  \label{eq48} \ee

$\bullet$ Particles with dispersion relation $\ep= \sum_j A_j n_j^{\gam_j} $, $n_j >0$ :

\be N_{\!<}(\ep)= {\ep^\ex \over \Gamma\left(1 + \ex \right)} \prod_j { \Gamma\left(1+ {1 / \gam_j}\right)
 \over A_j^{1/\gam_j}}
  \ee
 with $\ex= d_c + d_m/2$. The previous equations (\ref{eq44},\ref{eq45},\ref{eq46},\ref{eq48})  may be obtained from this formula with

$\gam_j=1$, \ \  $A_j=\hbar \pi c_j/L_j$ or $A_j=\hbar \omega_j$

$\gam_j=2$, \ \  $A_j=(\hbar^2 \pi^2 /(2 m_j L_j^2)$, 
\medskip

\noindent
and the \IDOS may be summarized in the form:

\be N_{\!<}(\ep)= {\ep^\ex \over   \Gamma\left( 1+  \ex  \right)} \prod_{j=1}^{d_c} \left({L_j\over \pi \hbar  c_j}\right)
 \prod_{j=1}^{d_m} \left(\sqrt{2 \pi m \over h^2} L_j \right) \label{gene1}
  \ee
 for a massless spectrum in $d_c$ directions and in the form:
  \be N_{\!<}(\ep)= {\ep^\ex \over   \Gamma\left(1+ \ex  \right)} \prod_{j=1}^{d_c} \left({1 \over  \hbar  \omega_j}\right)
 \prod_{j=1}^{d_m} \left(\sqrt{2 \pi m \over h^2} L_j \right)  \ . \label{gene2}
  \ee
  for a harmonic spectrum in $d_c$ directions.
\medskip

$\bullet$ Particles with dispersion relation $\ep= \sqrt{\sum_j (A_j n_j)^{2 \gam_j} }$, $n_j >0$:

\be N_{\!<}(\ep)= {\ep^\ex \over \Gamma\left(1 + \ex /2 \right)} \prod_j { \Gamma\left(1+ {1 / (2 \gam_j)}\right)
 \over A_j^{1/\gam_j}} \ee
With the above expressions of $A_j$, the \IDOS can be summarized as:
\begin{align}
 N_{\!<}(\ep)&= {\ep^\ex \over   \Gamma\left( 1+  \ex/2  \right)} \Gamma(3/2)^{d_c}  \Gamma(5/4)^{d_m}
  \nonumber \\
 & \times \prod_{j=1}^{d_c} \left({L_j\over \pi \hbar  c_j}\right)
 \prod_{j=1}^{d_m} \left(\sqrt{2 \pi m \over h^2} L_j \right) \label{gene1}
  \end{align}
   which is the same as for the dispersion $\ep= \sum_j A_j n_j^{\gam_j}$, within different numerical factors.

\subsection{Bosons and Fermions with zero $\mu$}

Here we collect the exact thermodynamic quantities for the saturated Bose gas and the compensated Fermi gas ($\mu=0$).
We start from a power law for the \IDOS, $N_{\!<}(\ep)=   B \ep^\ex $. The total energy is
\be U(T) = B \ex  \int_0^\infty {\ep^\ex \over e^{\gam \ep - \alpha} \pm 1} d\ep \ . \ee

For the saturated Bose gas,   the number of particles and the total energy read:\cite{pethick2002}

\begin{align} N(T)&=   B  \Gamma(1+\ex) \zeta(\ex) \,  T^{\ex}   \nonumber \\
 U(T)&=  \ex {\zeta(\ex+1) \over \zeta(\ex)} N(T) k_B T \ .
 \end{align}

For the compensated Fermi gas, we have

\begin{align} N(T)&=    B  \Gamma(1+\ex)(1 - 2^{1-\ex}) \zeta(\ex) \,  T^{\ex}  \nonumber \\
 U(T) &= \ex  {2^\ex-1 \over 2^\ex-2} {\zeta(\ex+1) \over \zeta(\ex)} N(T) k_B T \ .
\end{align}

\section*{acknowledgments}


The author acknowledges useful comments form J. Dalibard, J.-N. Fuchs and M.-O. Goerbig.


\begin{thebibliography}{99}
\bibitem{remark2} Also known as the \SB law is the expression for the emitted power per unit area ${\cal P}(T)= {c \over 4} u(T) \equiv \sigma T^4$ where $\sigma$ is named the \SB constant.
\bibitem{Fgf} In quantum theory, these functions are known to be $F(\alpha)= g_{{\red 1+ \kappa}}(e^\alpha)=\text{Li}_{1+ \kappa}(e^\alpha) $ for Bosons and
 $F(\alpha)= f_{{\red 1+ \kappa}}(e^\alpha)=-\text{Li}_{1+ \kappa}(-e^\alpha) $ for Fermions, where $\text{Li}_\nu(z)$ is the polylogarithm function.
\end{thebibliography}
\end{document}